\documentstyle[spie]{article} 
\input{psfig}   

\title{The Highest Redshift Radio Galaxies}

\author{Wil van Breugel\supit{a}
\skiplinehalf 
\supit{a}University of California - Lawrence Livermore National Laboratory \\ 
P.O. Box 808, Mailstop L-413, Livermore CA 94551, U.S.A.
\\
}

\authorinfo{Further author information: (Send correspondence to W.v.B.)\\W.v.B.: E-mail: wil@igpp.llnl.gov}

\begin{document} 
  \maketitle 

\begin{abstract}
At low redshifts powerful radio sources are uniquely associated with 
massive galaxies, and are thought to be powered by supermassive 
black holes. Modern 8m -- 10m telescopes may be used used to 
find their likely progenitors at very high redshifts to study 
their formation and evolution.

\end{abstract}




\newcommand\tna{{TN~J1338$-$1942~}}
\newcommand\tnb{{TN~J0924$-$2201~}}
\newcommand\cf{{c.f.,~}}
\newcommand\eg{{e.g.,~}}
\newcommand\etal{{et al.~}}
\newcommand\ie{{i.e.,~}}
\newcommand\kms{\ifmmode {\rm\,km\,s^{-1}}\else
    ${\rm\,km\,s^{-1}}$\fi}
\newcommand\lya{Ly$\alpha$}
\newcommand\Lya{Ly$\alpha$}
\newcommand\CIV{\hbox{C~$\rm IV$}~$\lambda$~1549}
\newcommand\HeII{\hbox{He~$\rm II$}~$\lambda$~1640}
\newcommand\OII{[\hbox{O~$\rm II$}]~$\lambda$~3727}
\newcommand\OIII{[\hbox{O~$\rm III$}]~$\lambda$~5007}
\newcommand\arcdeg{$^{\circ}$}
\newcommand\minpoint{\ifmmode \rlap.{^{\prime}}\else
    $\rlap.{^{\prime}}$\fi}
\newcommand\secpoint{\ifmmode \rlap.{^{\prime\prime}}\else
    $\rlap.{^{\prime}}$\fi}
\def\spose#1{\hbox to 0pt{#1\hss}}
\newcommand\simlt{\mathrel{\spose{\lower 3pt\hbox{$\mathchar"218$}}
     \raise 2.0pt\hbox{$\mathchar"13C$}}}
\newcommand\simgt{\mathrel{\spose{\lower 3pt\hbox{$\mathchar"218$}}
     \raise 2.0pt\hbox{$\mathchar"13E$}}}
\newcommand\nature{{Nature}}
\newcommand\aasup{{A\&AS}}
\newcommand\msun{\ifmmode {\rm\,M_\odot} \else
    ${\rm\,M_\odot}$\fi}
\newcommand\hzrgs{high redshift radio galaxies~}
\newcommand\hzrg{high redshift radio galaxy~}


\def\ebox{$\sqcup$\llap{$\sqcap$}}      
\def\sq{$\sqcup$\llap{$\sqcap$}}      
\def\ub{\underbar}
\def\hi{\noindent \hangindent=2.5em }
\def\ni{\noindent}
\def\header#1{\goodbreak\centerline{#1}}
\def\ls{\hfil\vskip 0.0truecm}
\def\spose#1{\hbox to 0pt{#1\hss}}
\def\iras{{\it IRAS }}
\def\IRAS{{\it IRAS }}
\def\km{{\rm\,km}}
\def\kms{\ifmmode {\rm\,km\,s^{-1}}\else
    ${\rm\,km\,s^{-1}}$\fi}
\def\kmsMpc{\ifmmode {\rm\,km\,s^{-1}\,Mpc^{-1}}\else
    ${\rm\,km\,s^{-1}\,Mpc^{-1}}$\fi}
\def\kpc{{\rm\,kpc}}
\def\mpc{{\rm\,Mpc}}
\def\msun{\ifmmode {\rm\,M_\odot}\else ${\rm\,M_\odot}$\fi}
\def\Msun{\ifmmode {\rm\,M_\odot}\else ${\rm\,M_\odot}$\fi}
\def\lsun{\ifmmode {\rm\,L_\odot}\else ${\rm\,L_\odot}$\fi}
\def\Lsun{\ifmmode {\rm\,L_\odot}\else ${\rm\,L_\odot}$\fi}
\def\rsun{\ifmmode {\rm\,R_\odot}\else ${\rm\,R_\odot}$\fi}
\def\Rsun{\ifmmode {\rm\,R_\odot}\else ${\rm\,R_\odot}$\fi}
\def\pc{{\rm\,pc}}
\def\cm{{\rm\,cm}}
\def\cm3{\ifmmode {\rm\,cm^{-3}}\else ${\rm\,cm^{-3}}$\fi}
\def\yr{{\rm\,yr}}
\def\gyr{{\rm\,Gyr}}
\def\Gyr{{\rm\,Gyr}}
\def\au{{\rm\,AU}}
\def\gm{{\rm\,g}}
\def\ergps{\ifmmode {\rm\,erg\,s^{-1}}\else ${\rm\,erg\,s^{-1}}$\fi}
\def\ergpscm2{\ifmmode {\rm\,erg\,s^{-1}\,cm^{-2}}\else
    ${\rm\,erg\,s^{-1}\,cm^{-2}}$\fi}
\def\K{{\rm\,K}}
\def\cgs{{\rm\,c.g.s.}}
\def\magn{{\rm\,mag}}
\def\cf{{\it cf.~}\ }
\def\eg{{\it e.g.}}
\def\dg{\mbox{$^\circ$}}        
\def\deg{\ifmmode {^{\circ}}\else {$^\circ$}\fi}
\def\degr{\ifmmode {^{\circ}}\else {$^\circ$}\fi}
\def\degs{\ifmmode {^{\circ}}\else {$^\circ$}\fi}
\def\degspt{$^\circ_\cdot$}
\def\etal{{\it et al.~}}
\def\Hb{{\rm\,H-$\beta$~}}
\def\hMpc{h^{-1}{\rm Mpc}}
\def\h3Mpc{h^{-3}{\rm Mpc}^3}
\def\Ho{\ifmmode {\rm\,H_\circ}\else ${\rm\,H_\circ}$\fi}
\def\hnot{\ifmmode {\rm\,H_\circ}\else ${\rm\,H_\circ}$\fi}
\def\h0{\ifmmode {\rm\,H_\circ}\else ${\rm\,H_\circ}$\fi}
\def\hnotunit{\ifmmode {\rm\,km\,s^{-1}\,Mpc^{-1}}\else
    ${\rm\,km\,s^{-1}\,Mpc^{-1}}$\fi}
\def\lya{{\rm\,Ly-$\alpha$~}}
\def\Lya{{\rm\,Ly-$\alpha$~}}
\def\qnot{\ifmmode {\rm\,q_\circ}\else ${\rm q_\circ}$\fi}
\def\q0{\ifmmode {\rm\,q_\circ}\else ${\rm q_\circ}$\fi}
\def\ie{{\it i.e.}}
\def\rms{\mbox{\it rms}}
\def\rmsub#1{\mbox{$_{\rm #1}$}}    
\def\ten#1{\times 10^{#1}}
\def\veps{\varepsilon}
\def\vs{{\it versus} }
\def\wth{\mbox{$w(\theta)$}}
\def\xc{cross-correlation }
\def\arcsec{\ifmmode {^{\prime\prime}~}\else $^{\prime\prime}~$\fi}
\def\asec{\ifmmode {^{\prime\prime}}\else $^{\prime\prime}$\fi}
\def\arcmin{\ifmmode {^{\prime}}\else $^{\prime}$\fi}
\def\amin{\ifmmode {^{\prime}}\else $^{\prime}$\fi}
\def\hr{$^{h}$}
\def\min{$^{m}$}
\def\mins{$'$\ }
\def\m{$^m$\ }
\def\mpt{$^m_\cdot$}
\def\hpt{$^h_\cdot$}
\def\sec{$^{s}$}
\def\s{$^s$\ }
\def\spt{$^s_\cdot$}
\def\secper{\ifmmode \rlap.{^{s}}\else $\rlap{.}{^{s}} $\fi}
\def\minper{\ifmmode \rlap.{^{m}}\else $\rlap{.}{^m} $\fi}
\def\magper{\ifmmode \rlap.{^{m}}\else $\rlap{.}{^m} $\fi}
\def\arcsper{\ifmmode \rlap.{^{\prime\prime}}\else
    $\rlap.{^{\prime\prime}}$\fi}
\def\arcmper{\ifmmode \rlap.{^{\prime}}\else
    $\rlap.{^{\prime}}$\fi}
\def\spose#1{\hbox to 0pt{#1\hss}}
\def\simlt{\mathrel{\spose{\lower 3pt\hbox{$\mathchar"218$}}
     \raise 2.0pt\hbox{$\mathchar"13C$}}}
\def\simgt{\mathrel{\spose{\lower 3pt\hbox{$\mathchar"218$}}
     \raise 2.0pt\hbox{$\mathchar"13E$}}}
\def\ltsima{$\; \buildrel < \over \sim \;$}
\def\gtsima{$\; \buildrel > \over \sim \;$}

\def\refindent{\par\noindent\parskip=2pt\hangindent=3pc\hangafter=1 }

%

\def\refp#1#2#3#4{\refindent{#1,} {#2}, #3, #4}
\def\refb#1#2#3{\refindent{#1}{ {#2}, }{#3}}
\def\refx#1{\refindent{#1}}


\section{Why the highest redshift radio galaxies are interesting}
\label{sect:why}  

Within standard Cold Dark Matter scenarios the formation of galaxies is
a hierarchical and biased process. Large galaxies are thought to grow
through the merging of smaller systems, and the most massive objects form
in over--dense regions, which will eventually evolve into the clusters
of galaxies seen today (\eg\ Ref.~\citenum{White97}).  It has also been suggested that
the first massive black holes may grow in similar hierarchical fashion
together with their parent galaxies (\eg\ Ref.~\citenum{Kauffmann00}) 
or, because of time scale constraints, may precede galaxy formation
and be primordial (\eg\ Ref.~\citenum{Loeb93}). It is therefore of great interest
to find the progenitors of the most massive galaxies and their AGN
(active massive black holes) at the highest possible redshifts and to
study their properties and cosmological evolution.

Radio sources are convenient beacons for pinpointing massive
elliptical galaxies, at least up to redshifts $z\sim 1$ (Ref.~\citenum{Lilly84};
Ref.~\citenum{Best98}). The near--infrared `Hubble' $K-z$
relation for such galaxies appears to hold up to $z= 5.2$, despite
large K--correction effects and morphological changes (Ref.~\citenum{WvB99b};
Fig.~\ref{fig:kz}). This suggests that radio sources may be used to
find massive galaxies and their likely progenitors out to very high
redshift through near--IR identification.

While optical, `color--dropout' techniques have been successfully used
to find large numbers of 'normal' young galaxies (without dominant
AGN) at redshifts even surpassing those of quasars and radio
galaxies (Ref.~\citenum{Weymann98}), the radio and near--infrared selection
technique has the additional advantage that it is unbiased with
respect to the amount of dust extinction. High redshift radio galaxies
are therefore also important laboratories for studying the
large amounts of dust (\eg\ Ref.~\citenum{Ivison98}) and molecular gas
(Ref.~\citenum{Papadopoulos00}), which are observed to accompany the
formation of the first forming massive galaxies.  Indeed, a
significant part of the scientific rationale for building future large
mm-arrays is based on the expectation that to understand galaxy
formation will ultimately require understanding their cold gas and
dusty environments.

Finally, it has been claimed that the (co--moving) space densities of
the most powerful radio galaxies and quasars were much higher near $z
\sim 2$, but that they drop off precipitously at even higher redshifts
(Ref.~\citenum{Dunlop90}; Ref.~\citenum{Shaver96}). However, using recently
completed studies of moderately faint radio galaxies (Ref.~\citenum{Jarvis99}) 
it has been argued that here is {\it no} such evidence for a redshift cut--off
and that these previous results have been biased due to unknown radio
K--correction, and thus radio spectral index, trends and associated
selection effects. 

\section{How to find the Highest Redshift Radio Galaxies}
\label{sect:find}

The near--infrared `Hubble' $K-z$ relation for radio galaxies (Fig.~\ref{fig:kz})
provides a convenient tool for finding radio galaxies at ever larger
redshifts. This was shown convincingly for the first time by Lilly 
(Ref.~\citenum{Lilly88}) 
who found that one of the faintest near-IR radio source identifications
in a complete, flux limited sample of $\sim 70$ objects, B3 0924+34,
was a redshift $z = 3.395$.

Unfortunately in complete, flux--limited samples the vast majority of
the sources will be relatively nearby, or at only modest redshifts. Lilly
(Ref.~\citenum{Lilly88}) in his survey found only 1 / 70 radio galaxies
at $z > 3$, and McCarthy \etal in a similar but $\sim 7$ times larger
survey also only found one (Ref.~\citenum{McCarthy96}). However, one can
pre-select very good HzRG candidates from the radio catalogs, before
even going to the telescope, by choosing sources with ultra--steep radio
spectra or `red radio color'.  It is already known more than 20 years
that the identification fraction of radio sources on the POSS plates
decreases with increasing spectral index (Ref.~\citenum{Tielens79}). One
had to wait for the much more sensitive CCD detectors before further
progress in identifying ultra-steep spectrum (USS) sources could be made.
One of the first HzRGs which was then found, using the Kitt Peak 4m,
was the radio galaxy 4C41.17, at $z = 3.800$ (Ref.~\citenum{Chambers90}). This
source was the record holder for many years, until it was by-passed,
using the same USS method, by 8C 1435+635 ($z = 4.25$; Ref.~\citenum{Lacy94})
and 6C 0140+326 ($z = 4.41$; Ref.~\citenum{Rawlings96}).

\begin{figure}[h]
\centering
\hspace{-0.5cm}
\begin{minipage}{8cm}
\centerline{\psfig{file=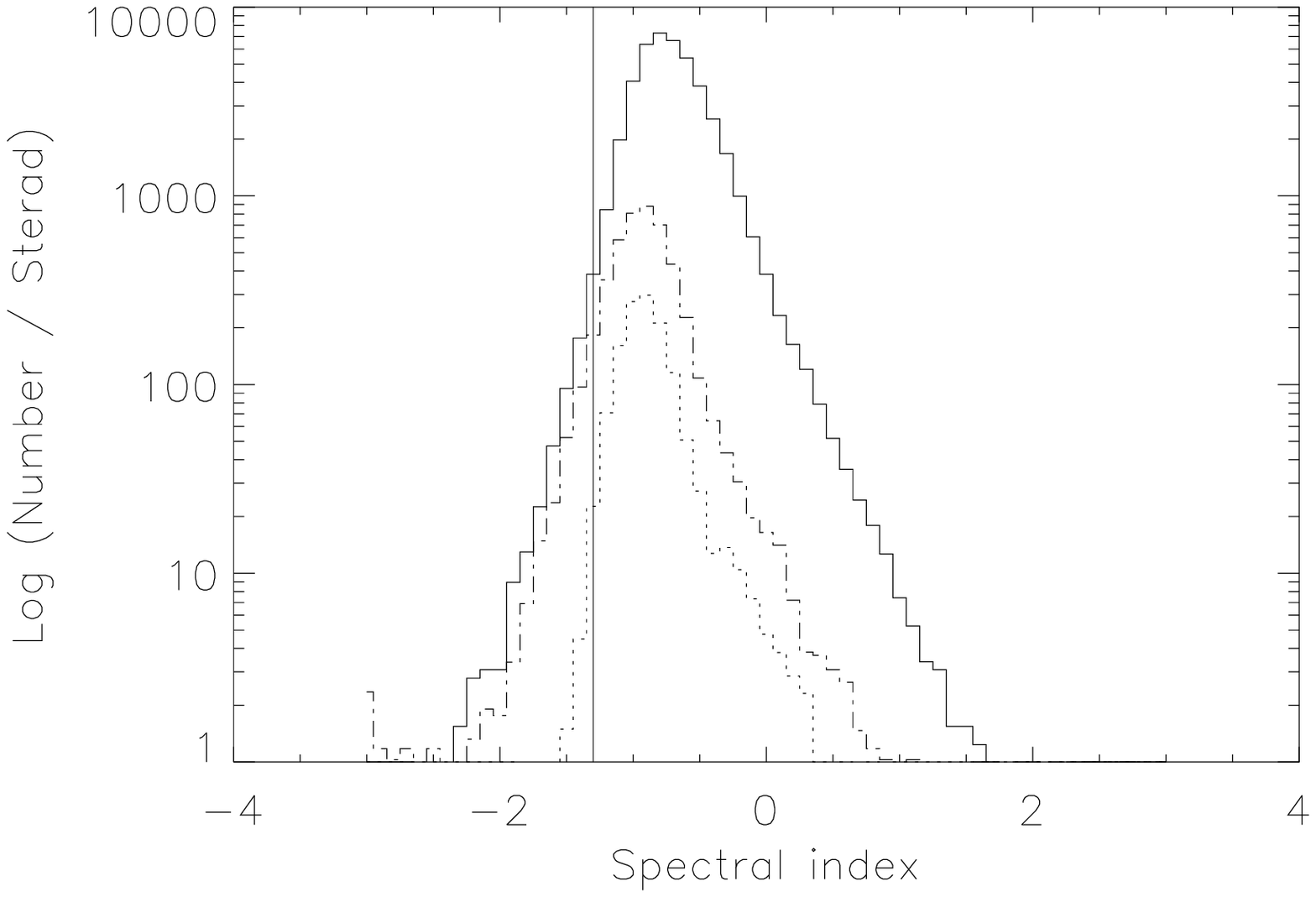,width=9cm}}
\caption{Logarithmic spectral index distribution for WENSS--NVSS (full line),
Texas--NVSS (dot--dash line) and MRC--PMN (dotted line). The vertical line indicates
the --1.3 cutoff used in our spectral index selection.
\label{fig:ussflux}
}
\end{minipage}
\hspace{0.5cm}
\begin{minipage}{8cm}
\centerline{\psfig{file=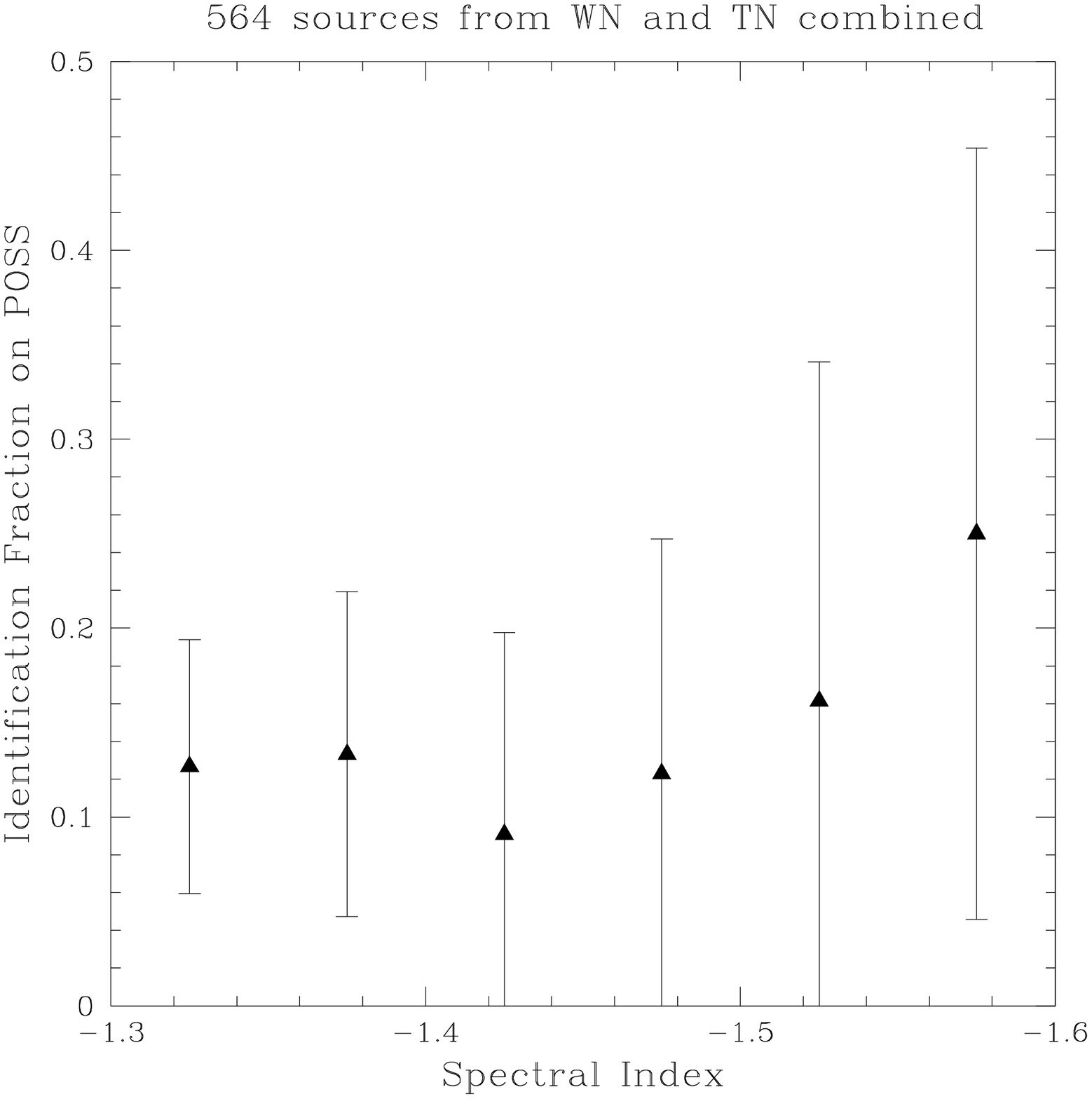,width=7cm}}
\caption{Identification fraction on the POSS as a function
of spectral index for the combined WN and TN sample.
\label{fig:ussposs}
}
\end{minipage}
\end{figure}

Together with graduate student C. De Breuck and colleagues at Leiden
Observatory we therefore defined the `ultimate' USS source sample by using
several new, large radio surveys (De Breuck \etal 2000$a$ [astro-ph/0002297]).  The sample
consists of 669 sources with extremely steep radio continuum spectra
($\alpha \le -1.3$; Table 1; Fig.~\ref{fig:ussflux}), at 10 -- 100 times lower flux
density limits than has been possible before (Ref.~\citenum{Chambers96};
Ref.~\citenum{Rottgering94}; Ref.~\citenum{Blundell98}).  To identify these
sources we first looked at the POSS and found that approximately $\sim$
15\% of the sources could be identified, usually with moderately bright
galaxies in galaxy clusters.  This identification fraction appears to be
independent of spectral index (Fig.~\ref{fig:ussposs}), at least for $\alpha \le -1.3$,
in support of the idea that these are mostly foreground objects.

\begin{table}[h]
\centering
\caption{USS samples$^a$}
\begin{tabular}{lcccc}
\hline
\hline
Sample & Density & Spectral Index & Flux Limit & \# of Sources \\
 & sr$^{-1}$ & ($S\nu \sim \nu^\alpha$) & mJy & \\
\hline
\hline
WN & 151 & $\alpha_{325}^{1400} \le -1.30$ & $S_{1400} >$ 10 & 343 \\
TN & 48$^b$ & $\alpha_{365}^{1400} \le -1.30$ & $S_{1400} >$ 10 & 268 \\
MP & 26 & $\alpha_{408}^{4800} \le -1.20$ & S$_{408} > 700$; S$_{4850} > 35$ & 58 \\
\hline
\end{tabular}
\begin{tabular}{p{11cm}}
\quad $^a$ See De Breuck \etal 2000$a$ (astro-ph/0002297) for catalogs used and other details. \\
\quad $^b$ Due to the characteristics of the Texas survey, the TN sample
is only $\sim 30\%$ complete.
\end{tabular}
\end{table}

The USS selection proved to be extremely efficient. Attempts to obtain
optical identifications of USS sources using 3m--4m--class telescopes
($R \simlt 24$) were largely unsuccessful. Also near--IR imaging would
be very difficult, given the typically expected $R - K \sim 4$ values of
\hzrgs . We therefore decided to entirely skip the optical identification
program at Lick Observatory 
and go straight to near--IR imaging at the Keck I telescope. 

\section{Morphological evolution of the highest redshift radio galaxies}
\label{sect:near-IR}

When we started our near--IR imaging program at Keck our first order
of business was to observe \hzrgs with known redshifts $z > 1.9$ to
investigate their morphological evolution and to obtain more accurate
photometry to study the Hubble $K-z$ diagram at the highest redshifts.
We obtained near--IR images of 15 HzRGs with $1.9 < z < 4.4$ with the
Near Infrared Camera (NIRC, Ref.~\citenum{Matthews94}) at the Keck I telescope.
The images show that there is strong morphological evolution at {\it
rest--frame optical} ($\lambda_{\rm rest} > 4000$\AA) wavelengths
(Ref.~\citenum{WvB98}; Fig.~\ref{fig:nirc3c257}).  At the highest redshifts, $z > 3$,
the rest--frame visual morphologies exhibit structure on at least two
different scales: relatively bright, compact components with typical
sizes of $\sim 1$ \arcsec ($\sim$10 \kpc) surrounded by large--scale ($\sim$ 50 \kpc)
diffuse emission. The brightest components are often aligned with the
radio sources, and their {\it individual} luminosities are $M_B \sim -20$
to $-22.$ For comparison, present--epoch L$_\star$ galaxies and, perhaps
more appropriately, ultraluminous infrared starburst galaxies, have,
on average, $M_B \sim -21.0$. The {\it total, integrated} rest--frame
B--band luminosities are $3 - 5$ magnitudes more luminous than present
epoch $L_\star$ galaxies.

At lower redshifts, $z < 3$, the rest--frame optical morphologies become
smaller, more centrally concentrated, and less aligned with the radio
structure.  Galaxy surface brightness profiles for the $z < 3$ HzRGs
are much steeper than those of at $z > 3$.  We attempted to fit the $z <
3$ surface brightness profiles with a de Vaucouleurs r$^{1/4}$ law and
with an exponential law, the forms commonly used to fit elliptical and
spiral galaxy profiles, respectively.  We demonstrate the fitting for
our best resolved object at $z < 3$, 3C 257 at $z = 2.474$ (Fig.~\ref{fig:nirc3c257}).
Within the limited dynamical range of the data, both functional forms
fit the observed profiles---neither is preferred.  Interestingly,
despite this strong morphological evolution the $K - z$ Hubble diagram
for the most luminous radio galaxies remains valid even at the highest
redshifts, where a large fraction of the K-band continuum is due to a
radio--aligned component.

Having established that the $K-z$ diagram for \hzrgs holds even at
the highest known redshifts we embarked on our identification program
of USS selected sources. Our typical method of observation would be
to begin with 16 x 1 minute exposures (1 minute consisting of 2 or 3
co--added frames), start a second 16 x 1 minute run while reducing the
first set of observations using DIMSUM.  (DIMSUM is the Deep Infrared
Mosaicing Software package, developed by P.\ Eisenhardt, M.\ Dickinson,
A.\ Stanford, and J.\ Ward, which is available as a contributed package
in IRAF.) If we could identify our target we would break off our second
observation, or, if the identification was faint, would let it finish and then go
on to the next target. This `on--line' way of observing turned out to
be very efficient and resulted in a 100\% identification rate with good
photometric magnitudes and has provided excellent \hzrg candidates using
the Hubble $K-z$ diagram (Fig.~\ref{fig:kz}; Fig.~\ref{fig:nircpanel}).
Often more than a dozen near--IR identifications could be obtained this
way in a single night.

\section{Spectroscopy of the highest redshift radio galaxies}
\label{sect:spec}

As with our near--IR imaging program, our first spectroscopic
observations, using the Low Resolution Imaging Spectrograph (LRIS,
Ref.~\citenum{Oke95}) were made of \hzrgs with known redshifts. The main
purpose, initially, was to determine the origin of the
radio--aligned optical / near--IR features using spectro--polarimetry.
As is now well--known the rest--frame optical continua of \hzrgs are
often clumpy and aligned with their associated radio sources 
(Ref.~\citenum{McCarthy87}; Ref.~\citenum{Chambers87}).
This suggested that there must be
a causal connection between their optical morphological appearance
and the collimated outflow and/or ionizing radiation from their AGN.
The most popular explanations for such an alignment effect are scattered
light from hidden or mis--aligned quasars, jet--induced star formation
or nebular continuum emission. Evidence for each of these processes
has been found.
In particular, $z \sim 1$
and $z \sim 2.5$ most \hzrgs are strongly polarized, indicating that
a large fraction of the optical continuum is due to scattered light
from hidden or mis--aligned quasars (Ref.~\citenum{Cimatti99}; 
Ref.~\citenum{Dey99};Ref.~\citenum{Vernet99}). However, deep spectropolarimetry
observations of two $z > 3.5$ radio galaxies (4C 41.17 at $z = 3.800$
and 6C J1908+722 at $z = 3.534$) show no polarized continua but instead
show evidende for absorption lines from young hot stars (Ref.~\citenum{Dey99}).
It suggests that at the highest redshifts radio
galaxy hosts are dominated by massive starbursts, possibly induced by
radio jets (Ref.~\citenum{Dey97};Ref.~\citenum{WvB99a})
and not by scattered light from their AGN.

Subsequently our spectroscopic observations focused on the newly
identified USS \hzrg candidates.  At the present time we have observed
and analyzed 34 USS \hzrgs with the following results.  Only 5 of the
sources have $z<2$, 8 have $2<z<3$, 9 have $3<z<4$ and 3 sources have
$z>4$, including one at $z > 5$. At least 3 sources were not detected
in optical continuum, despite $\sim$ 1 hr or longer integrations with
LRIS. All we know of these objects is that they are detected in the
near-IR at $K \sim 21$, and have a radio source identified with them. They
may be extremely obscured, or 
at record high redshifts, with \Lya redshifted to near-IR wavelengths
($z > 8$). Future observations with near--IR
spectrographs may tell.  We also found 6 sources with only
a continuum detection and no emission--lines. These were all extremely
compact USS sources, and may be moderately high redshift ($1 < z < 3$)
BL Lac objects, `emission--line free quasars' (\cf Ref.~\citenum{Fan99}), or
even pulsars (which typically have $\alpha_{radio} \sim -1.6$, 
Ref.~\citenum{Kaplan98}, 
and are faint optically).

The \hzrg spectra, when obtained with sufficient spectral resolution, show
nearly all very strong blueward asymmetries in the \lya emission lines
(Ref.~\citenum{Dey99}; Ref.~\citenum{vanOjik97}; De Breuck \etal 2000$b$
(in preparation); Fig.~\ref{fig:lya-abs}. This is almost certainly due
to the presence of cold gas (HI) and dust in the vicinity of the radio
galaxies, not just because of cosmological \lya `forest' absorption
in the foreground (although this will contribute as well). Spatially
resolved emission line regions show that this absorption can occur over
the entire region (up to $\sim$ 50 kpc; Fig.~\ref{fig:lya-abs}), and
is strongest in the smallest radio sources (Ref.~\citenum{vanOjik97}).
There is much additional evidence for the presence of large amounts
of cold gas in dust in \hzrgs . Many of the highest redshift radio
galaxies have been detected at sub--mm wavelengths, both in continuum
(\eg Ref.~\citenum{Ivison98}; Archibald \etal 2000 [astro-ph/0002083])
and molecular lines (\eg Ref.~\citenum{Papadopoulos00}).
These observations indicate total dust masses
of 10$^8$ -- 10$^9$ \msun, and star formation rates of more than 1000
\msun/yr. Thus \hzrgs indeed appear to be massive forming systems.

One object deserves special mention: 6C J1908+722 at $z = 3.53$
(Ref.~\citenum{Dey99}; Fig.~\ref{fig:balrag1}).  The source shows very broad
absorption lines in several of its UV resonant lines (CIV, SIV, NV,
\lya; Fig.~\ref{fig:balrag2}).  This was interpreted as being caused by
outflow, similar to the classical Broad Absorption Line quasars. However,
it is interesting to note that this Broad Absorption Line Radio Galaxy
(BALRAG) is hosted by a $\sim 1.5 \times 10^{13}$ \lsun Ultra Luminous
Infrared Galaxy with $\sim 1.5 \times 10^{8}$ \msun~ in dust, $\sim 5 \times
10^{10}$ \msun~ in molecular gas, and has an estimated star formation rate of
$\sim 1500$ \msun/yr (Ref.~\citenum{Papadopoulos00}). The observed velocity
range of the gas is large (530 \kms), and could be even larger: 
for another \hzrg, 4C60.07 at $z = 3.788$, Papadopoulos \etal find that
the molecular gas is distributed over at least two major components,
with a total velocity range $> 1000$ \kms. 

Thus it could very well be that the broad, rest--frame UV, absorption
lines in 6C J1908+722 may be due to absorption within the parent
galaxy. The large BAL velocity range (Fig.~\ref{fig:balrag2}) could
then be caused by a number of cold gas components in the foreground to
6C J1908+722, and which could be falling in or merging with the galaxy.
In that case one would expect that the BAL system would be resolved at
higher spectral resolution and observations at Keck to test this are planned.

\section{The Highest Redshift Radio Galaxies}
\label{sect:highest}

\subsection{TN~J1338$-$1942 at $z = 4.11$}
\label{sect:tn1338}

The first $z > 4$ USS radio galaxy discovered by us was
TN~J1338$-$1942. The initial identification was made with the ESO
3.6m at R--band, and subsequent spectroscopy with that same telescope
showed that the radio galaxy has a redshift of $z=4.11 \pm 0.02$, based
on a strong detection of \lya, and weak confirming \CIV\ and \HeII\
(Ref.~\citenum{DeBreuck99}).

\begin{figure} [h]
\centering
\hspace{-0.5cm}
\begin{minipage}{8cm}
\centerline{\psfig{file=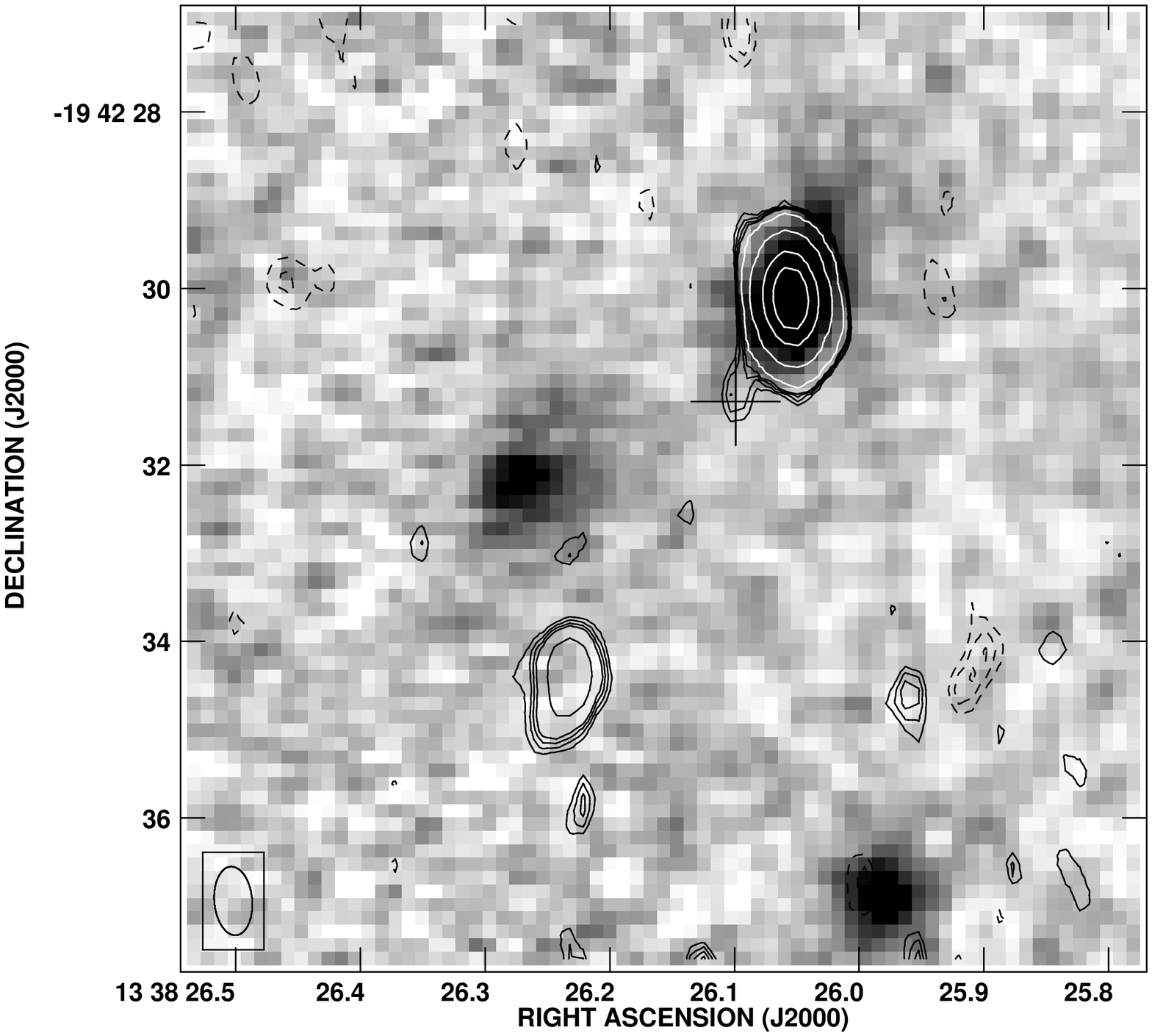,width=8cm}}
\caption{4.85~GHz VLA radio contours overlaid on a Keck $K-$band
image of \tna.
\label{fig:tn1338-nirc}
}
\end{minipage}
\hspace{0.5cm}
\begin{minipage}{8cm}
\centerline{\psfig{file=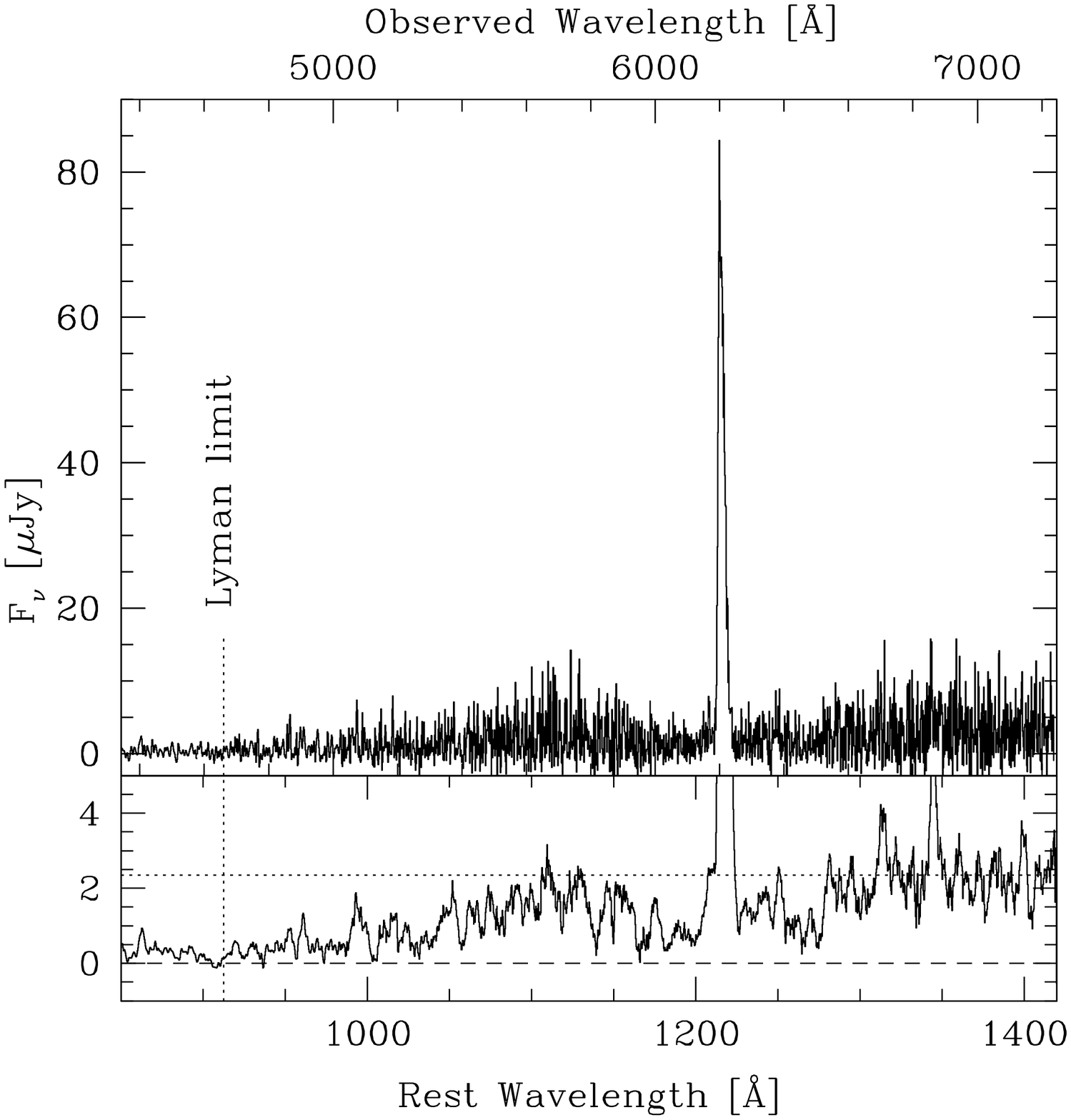,width=7cm}}
\vspace{1.5mm}
\caption{VLT spectrum of \tna\ at $z = 4.11$.
\label{fig:tn1338-vlt}
}
\end{minipage}
\end{figure}

Subsequently we obtained a deep K--band image (rest--frame B--band)
at Keck, shown in Fig.~\ref{fig:tn1338-nirc} overlaid with a VLA
radio image (Ref.~\citenum{DeBreuck99}).  The \Lya\ and rest--frame
optical emission appear co--spatial with the brightest radio hotspot
of this very asymmetric radio source.  Such asymmetric radio sources
are not uncommon, even in the local Universe, and are usuallly thought
to be due to strong interaction of one of its radio lobes with very
dense gas.  A similar asymmetric radio/optical/emission-line morphology
has also been seen in the $z = 3.800$ radio galaxy 4C41.17, where it
has been interpreted as being caused by jet-induced star formation
(Ref.~\citenum{Dey97};Ref.~\citenum{WvB99a}; Bicknell \etal 2000
[astro-ph/9909218]).

With the Keck K-band the identification and astrometry for \tna secured
we next obtained a high signal--to--noise, medium resolution (5.5~\AA\
FWHM) spectrum using the VLT Antu telescope (Ref.~\citenum{DeBreuck99};
Fig.~\ref{fig:tn1338-vlt}). The spectrum of \tna is dominated by a bright
\Lya\ line ($W_{\rm Ly\alpha}^{\rm rest}$ = 210\AA) which shows deep and
broad ($\sim 1400$ \kms) blue--ward absorption, and relatively bright
($F_{1400} \sim 2 \mu$Jy) UV-continuum. In fact, at optical wavelengths,
\tna turned out to be the most luminous of its kind (Table 2).  If all
the UV continuum in \tna would be due to young O--B stars the implied SFR,
based on the optical data alone and  
without correction for extinction, 
would be several hundred \msun/yr, similar to 4C41.17. \tna might
be another example of a very HzRG in which jet-induced star formation
might occur.

The \Lya\ is spatially extended by $\sim$ 4\arcsec\ (30~kpc) and has
a spectral profile that is very asymmetric with a deficit towards
the blue. This blue-ward asymmetry is probably due
to absorption of the \Lya\ photons by cold gas in a turbulent halo
surrounding the radio galaxy.  Using a simple model, and fitting the
\Lya\ profile with a Gaussian emission function and a single Voigt
absorption function, De Breuck \etal estimate that the neutral hydrogen
column density must be in the range $3.5 - 13 \times 10^{19}$ cm$^{-2}$,
and a total mass of $2 - 10 \times 10^7 M_{\odot}$.

The bright optical continuum and high S/N data also allowed a measurement
of the Lya\ forest continuum break (\Lya\ 'discontinuity', $D_A$), and
the Lyman limit.  The measured value, $D_A=0.37 \pm 0.1$, is $\sim
0.2$ lower than the values found for quasars at comparable redshifts.
This might perhaps be due due to a bias towards large $D_A$ introduced
in high--redshift quasar samples that are selected on the basis of large
color gradients. The true space density of optically selected quasars,
-- and Lyman break galaxies --, may have been underestimated and the
average HI column density along cosmological lines of sight might have
been overestimated. Because of their radio-based, non--color selection,
$z>4$ radio galaxies may be excellent objects for investigating $D_A$
statistics.

\subsection{TN~J0924$-$2201 at $z = 5.19$}
\label{sect:tn0924}

\begin{figure}[h]
\centering
\hspace{-0.5cm}
\begin{minipage}{8cm}
\caption{Keck/NIRC $K$-band image of \tnb, with radio contours superposed.
\label{fig:tn0924-nirc}
}
\end{minipage}
\hspace{0.5cm}
\begin{minipage}{8cm}
\centerline{\psfig{file=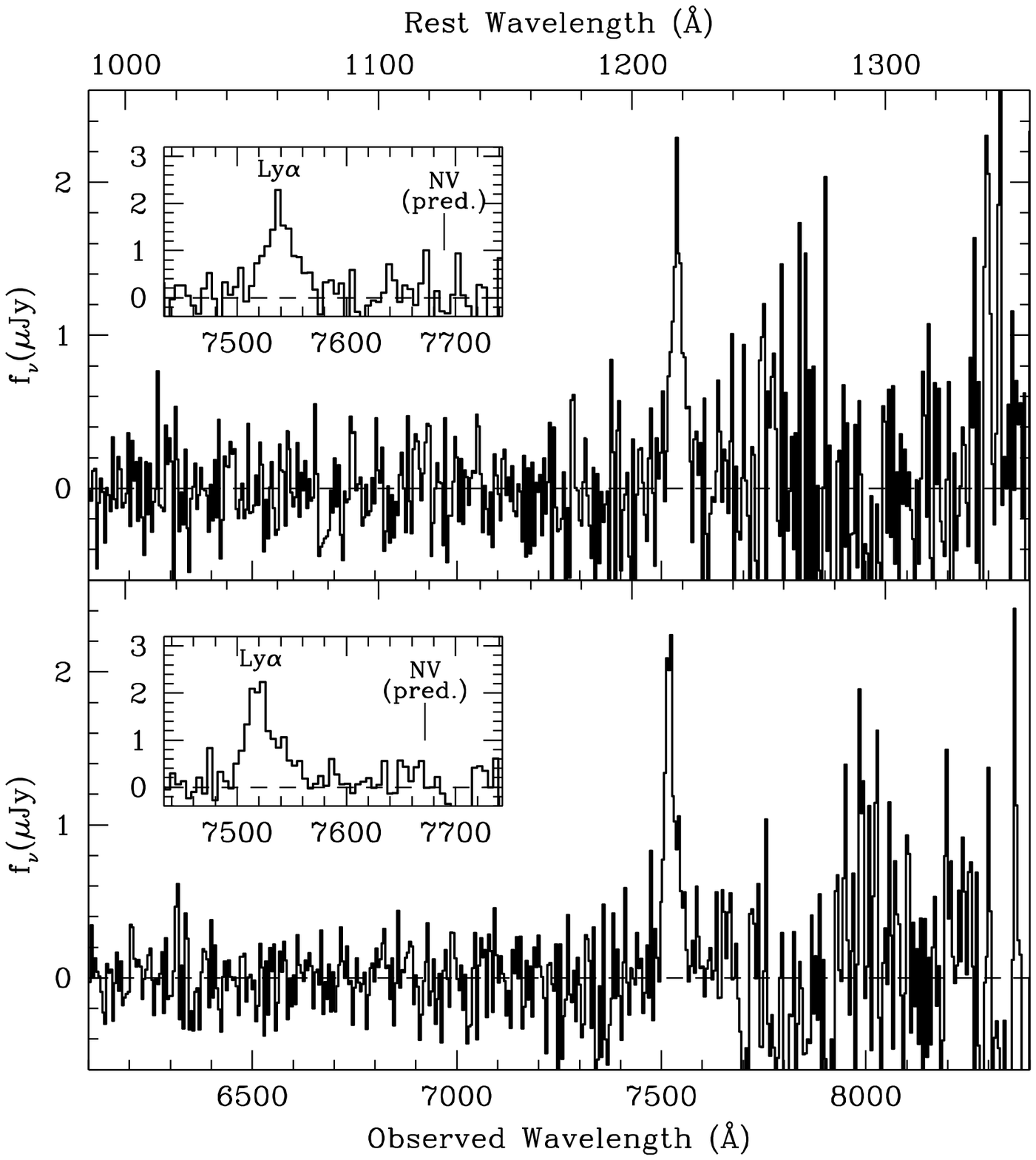,width=8cm}}
\vspace{-5mm}
\caption{Keck spectra of \tnb at $z = 5.19$ on two different nights.
\label{fig:tn0924-spec}
}
\end{minipage}
\end{figure}

TN~J0924$-$2201 is one of the steepest spectrum sources in our USS
sample ($\alpha_{\rm 365 MHz}^{\rm 1.4 GHz} = -1.63$) and therefore
was one of our primary targets for near--IR identification.  A deep
K--band image at Keck showed indeed a very faint ($K = 21.3 \pm 0.3$),
multi--component object at the position of the small (1.2\arcsec) radio
source (Fig.~\ref{fig:tn0924-nirc}).  The expected redshift on the basis
of the $K-z$ diagram was $z > 5$, and spectroscopic observations at
Keck showed that this was indeed the case (Fig.~\ref{fig:tn0924-spec}),
based on a single emission line at $\lambda \sim 7530$ \AA\ which we
identified as \lya\ at $z = 5.19$ (Ref.~\citenum{WvB99b}; none of the
$z > 5$ galaxies have more than one line detection).

Among all radio selected \hzrgs \tnb\ is fairly typical in radio
luminosity, equivalent width and velocity width (Table~2). It does have
the steepest radio spectrum, consistent with the $\alpha - z$ relationship
for powerful radio galaxies (\eg\ Ref.~\citenum{Rottgering97}), and also has
the smallest linear size.  The latter may be evidence of its `inevitable
youthfulness' or a dense confining environment, neither of which would be
surprising because of its extreme redshift (Ref.~\citenum{Blundell99};
Ref.~\citenum{vanOjik97}).  Among the radio selected \hzrgs \tnb\ appears
underluminous in \lya, together with 8C~1435$+$63, which might be caused
by absorption in an exceptionally dense cold and dusty medium. Evidence for
cold gas and dust in several of the most distant \hzrgs has been found
from sub--mm continuum and CO--line observations (\eg Ref.~\citenum{Ivison98};
Ref.~\citenum{Papadopoulos00}).

The second highest redshift radio galaxy currently known listed in
Table 2 is VLA~J123642+621331 at $z = 4.42$ (Ref.~\citenum{Waddington99}).
This source was not USS selected and provides 
a view on the possible selection effects of our 
USS \hzrgs . The source is an asymmetric double and
although its radio luminosity is about a factor 1000 times lower than
that of its much more luminous brothers at similar redshifts, it is still
radio loud, with a radio luminosity close to the 
FRI / FRII break at 408 MHz ($P_{408} \sim 3.2 \times 10^{33}$
erg s$^{-1}$ {Hz}$^{-1}$).
Its radio spectrum is steep ($\alpha_{8.4GHz}^{1.4GHz}
\sim -1.0$, using the flux densities given by Waddington \etal), but
not as steep as our USS selected \hzrgs, and the \lya\ luminosity is a
factor 5 -- 10 times less. Apart from the luminosity these properties
are not hugely different from expected on the basis of radio selection.
It suggests that less extreme steep spectrum selected samples
($\alpha < -1.0$) at much lower flux densities ($\simlt 1$ mJy) might
be used to find many more \hzrgs at very high 
redshifts, although with lower efficiency, we suspect, than USS
selected samples.

\begin{table}
\caption{Physical parameters of the highest redshift radio galaxies} \label{tbl-2}
\begin{center}
\small
\begin{tabular}{crrrrrrrc}
\hline
\hline
Name & $z$ & {$L_{\rm Ly\alpha}$ $^a$} & {$L_{\rm 365}$ $^a$}
& $\alpha_{\rm 365}^{\rm 1400}$ & $W_{\rm Ly\alpha}^{\rm
rest}$ & {$\Delta_{\rm Ly\alpha}$ $^a$} &
\multicolumn{1}{c}{Size $^a$} & {Ref. $^b$} \\
\hline
TN~J0924$-$2201 &  5.19 & 1.3 & 7.5 & $-$1.63 &  $>$115 & 1500 &  8 & WvB99 \\
VLA~J1236+6213 & 4.42 & 0.2 & 0.0035 & $-$0.96 &  $>$50 &  440 &    & Wad99 \\
6C~0140$+$326   &  4.41 &  16 & 1.3 & $-$1.15 &     700 & 1400 & 19 & DeB00 \\
8C~1435$+$63    &  4.25 & 3.2 & 11  & $-$1.31 &     670: & 1800 & 28 & Spin95 \\
TN~J1338$-$1942 &  4.11 &  25 & 2.3 & $-$1.31 &     200 & 1000 & 37 & DeB99 \\
4C~41.17        & 3.798 &  12 & 3.3 & $-$1.25 &     100 & 1400 & 99 & Dey97 \\
4C~60.07        &  3.79 &  16 & 4.1 & $-$1.48 &     150 & 2900: & 65 & R\"ot97 \\
\hline
\end{tabular}
\end{center}
\centering
\begin{tabular}{p{12cm}}
\quad $^a$ In units of $10^{43}$ erg s$^{-1}$ ($L_{\rm Ly\alpha}$),
$10^{36}$ erg s$^{-1}$ {Hz}$^{-1}$ ($L_{\rm 365}$),
restframe velocity width km $s^{-1}$, and linear size in kpc (respectively) \\
\quad $^b$ Most recent references quoted only:
WvB99 = Ref.~\citenum{WvB99b};
Wad99 = Ref.~\citenum{Waddington99};
DeB00 = De Breuck \etal 2000$b$ (in preparation);
Spin95 = Ref.~\citenum{Spinrad95};
DeB99 = Ref.~\citenum{DeBreuck99};
Dey97 = Ref.~\citenum{Dey97};
R\"ot97 = Ref.~\citenum{Rottgering97}.
\end{tabular}
\end{table}

\begin{figure}[h]
\centerline{\psfig{file=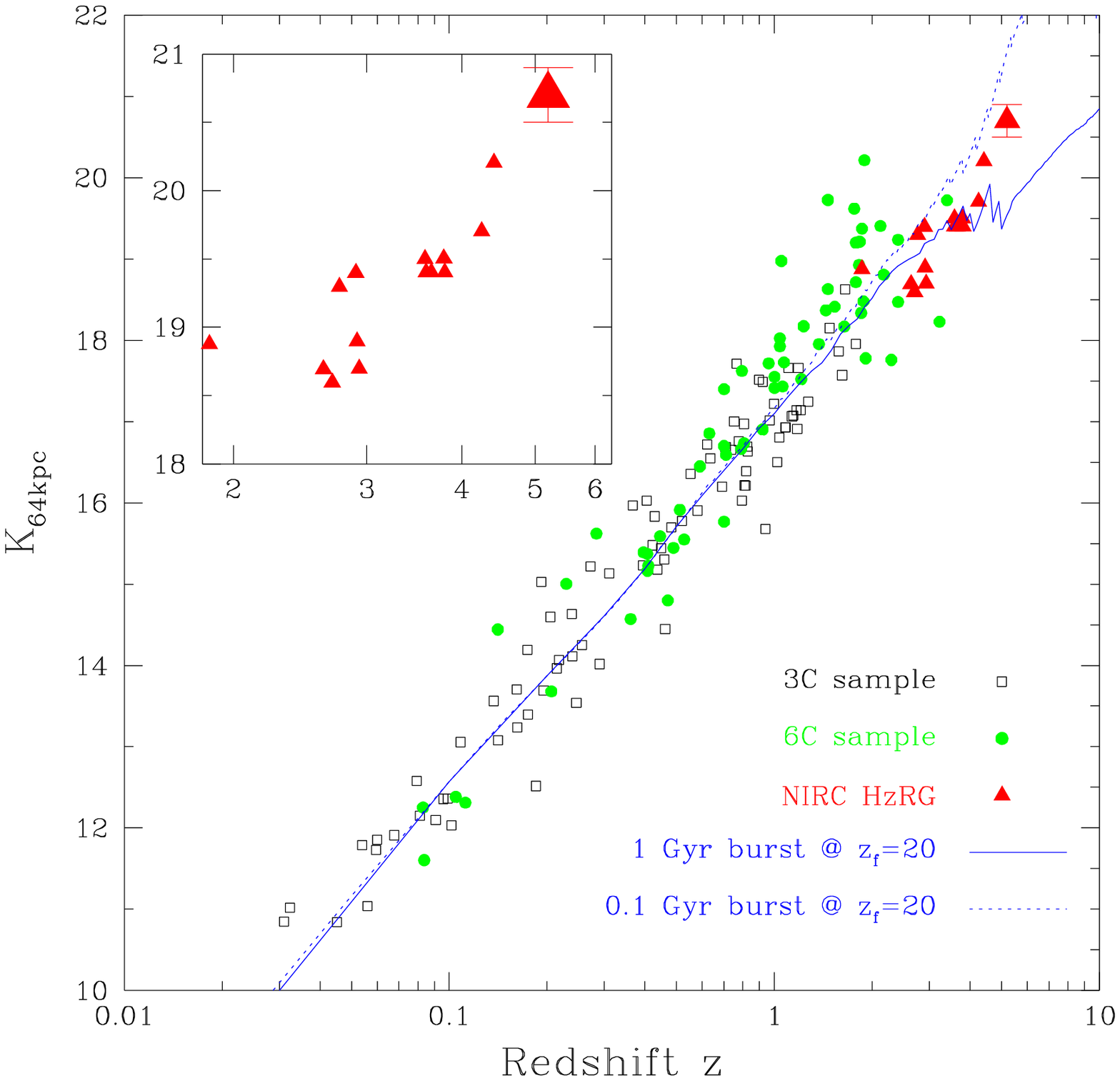,width=12cm}}
\caption{The Hubble $K-z$ diagram for
\hzrgs .  Filled triangles are Keck
measurements of \hzrgs from Ref.~14, the large triangle
is \tnb\ at $z = 5.19$, and all other photometry is from Ref.~15.
Two stellar evolution models
from Bruzual and Charlot (1999; priv. comm.)
normalized at $z<0.1$, are plotted, assuming parameters as shown.
\label{fig:kz}
}
\end{figure}

Our observations of \tnb\ extend the Hubble $K-z$ diagram for powerful
radio galaxies to $z = 5.19$, as shown in Fig.~\ref{fig:kz}.  Simple stellar
evolution models are shown for comparison.  Despite the enormous
$k$--correction effect (from $U_{\rm rest}$ at $z = 5.19$ to
$K_{\rm rest}$ at $z = 0$) and strong morphological evolution (from
radio--aligned to elliptical structures), the $K-z$ diagram remains a
powerful phenomenological tool for finding radio galaxies at extremely
hy redshifts. Deviations from the $K-z$ relationship may exist 
Ref.~\citenum{Eales97}; but see Ref.~\citenum{McCarthy99}), and scatter in the $K-z$ values
appears to increase with redshift. Part of this may be due to lack of 
S/N or contamination by strong line emission in some of the measurements.

The clumpy $U_{\rm rest}$ morphology resembles that of other \hzrgs 
(Ref.~\citenum{WvB99a}; Ref.~\citenum{Pentericci98}) and if it is dominated by
star light we derive a SFR of $\sim$200 M$_\odot$ yr$^{-1}$, without any
correction for extinction, which may be a factor of several.  \tnb\ may
be a massive, active galaxy in its formative stage, in which the SFR is
boosted by jet--induced star formation.
For comparison other, `normal' star forming
galaxies at $z > 5$ have 10 -- 30 times lower SFR ($\sim 6 - 20$ \msun/yr).

At the time of its discovery, December 1998, \tnb\ was the most
distant AGN known, surpassing even quasars for the first time since
their discovery 36 years ago.  The recent, serendipitous discovery of
a color--selected $z = 5.50$ quasar (Stern \etal 2000 [astro-ph/0002338]) returned the
record to optically selected quasars.  The presence of AGN at such early
epochs in the Universe ($< $1 Gyr in most cosmogonies) poses interesting
challenges to common theoretical wisdom, which assumes, at least for
radio loud AGN, that they are powered by massive (billion solar mass),
active black holes. The question how these can form so shortly after
the putative Big Bang may prove even more challenging then that of the
formation of galaxies (\eg Ref.~\citenum{Loeb93}).

\acknowledgments     
 
\begin{figure}[h]
\centering
\hspace{-0.5cm}
\begin{minipage}{8cm}
\centerline{\psfig{file=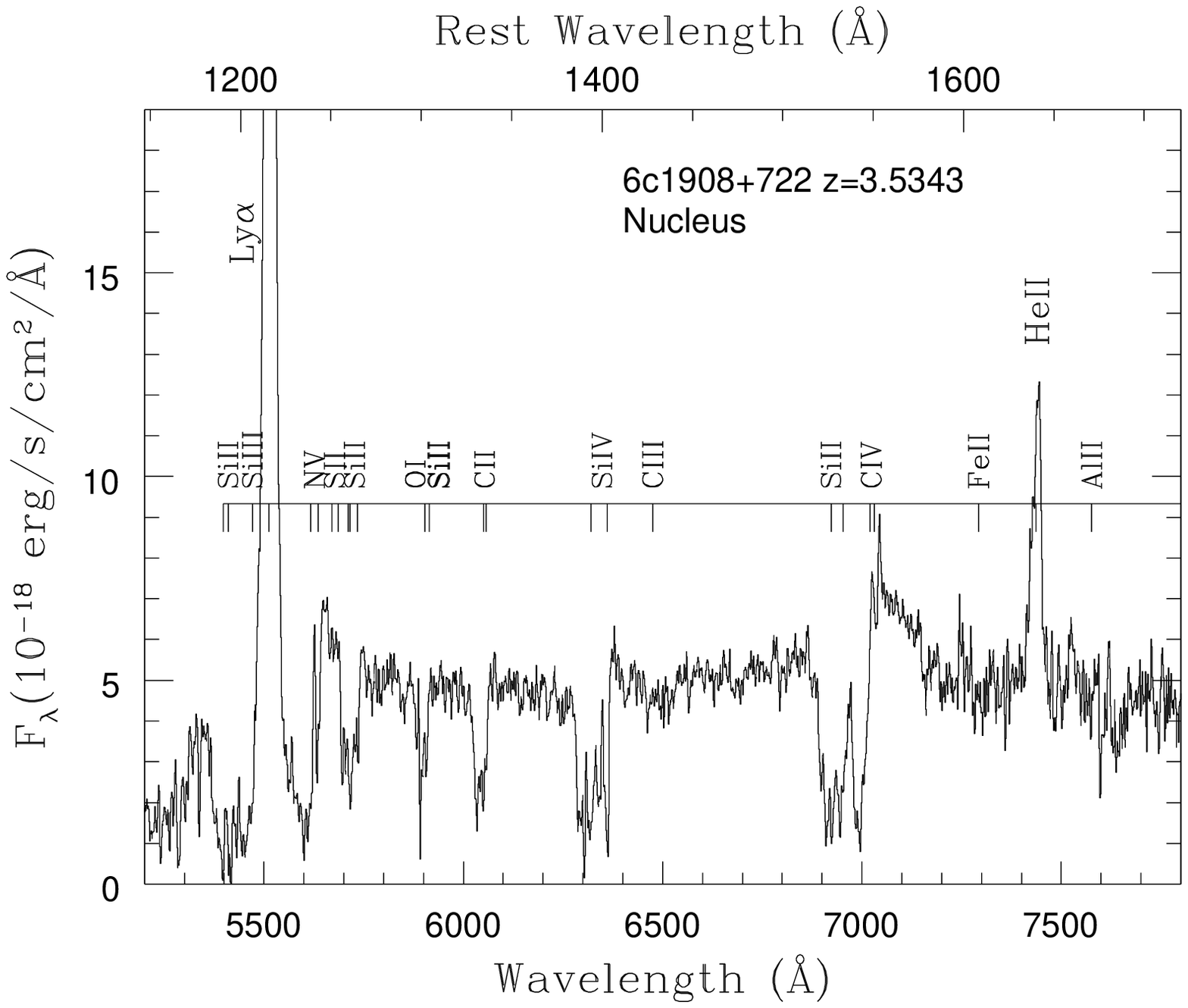,width=11cm}}
\vspace{-4mm}
\caption{Spectrum of the $z = 3.53$ BALRAG 6C~1908+722 (Figure from Ref.~29).
\label{fig:balrag1}
}
\end{minipage}
\hspace{0.5cm}
\begin{minipage}{8cm}
\centerline{\psfig{file=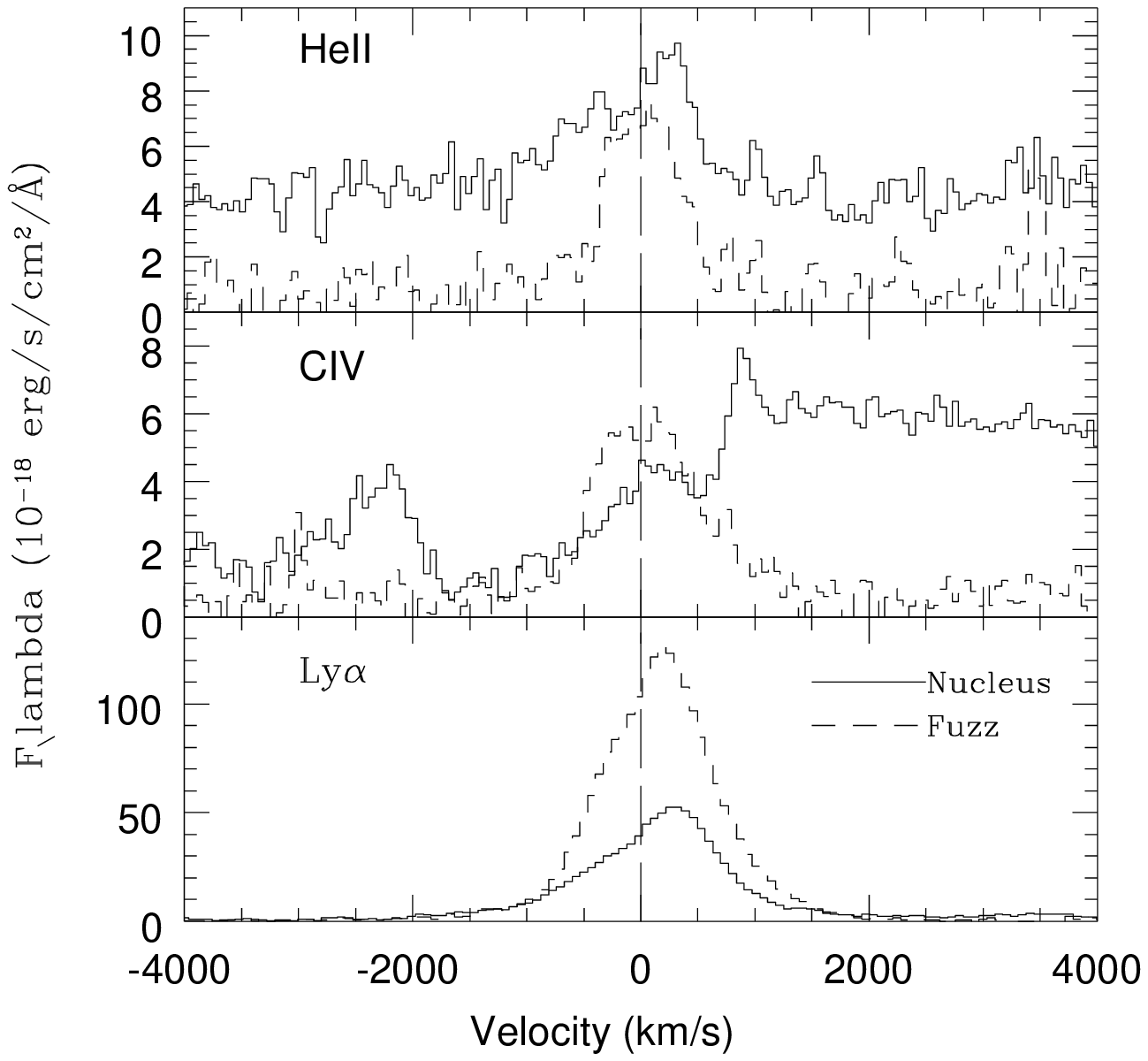,width=9.5cm}}
\vspace{-1.5cm}
\caption{The Ly$\alpha$, CIV and HeII line profiles. Note that the
profiles appear to be asymmetric both on and off the nucleus
(Figure from Ref.~29).
\label{fig:balrag2}
}
\end{minipage}
\end{figure}

WvB thanks his many collaborators for stimulating discussions and fun
observing runs. Special thanks to C. De Breuck, who has done much of the
work as part of his thesis research, and A. Dey for permission to use
Fig.~\ref{fig:lya-abs}, Fig.~\ref{fig:balrag1} and Fig.~\ref{fig:balrag2}.
The work by W.v.B.  at the University of California Lawrence Livermore
National Laboratory was performed under the auspices of the US Department
of Energy under contract W-7405-ENG-48.  W.v.B.\ also acknowledges
support from several NASA grants in support of \hzrg research with HST.


\bibliography{report}   
\bibliographystyle{spiebib}   


\vfill\eject


\vfill\eject


\vfill\eject


\hspace{2.0cm}

\noindent{This figure is available as 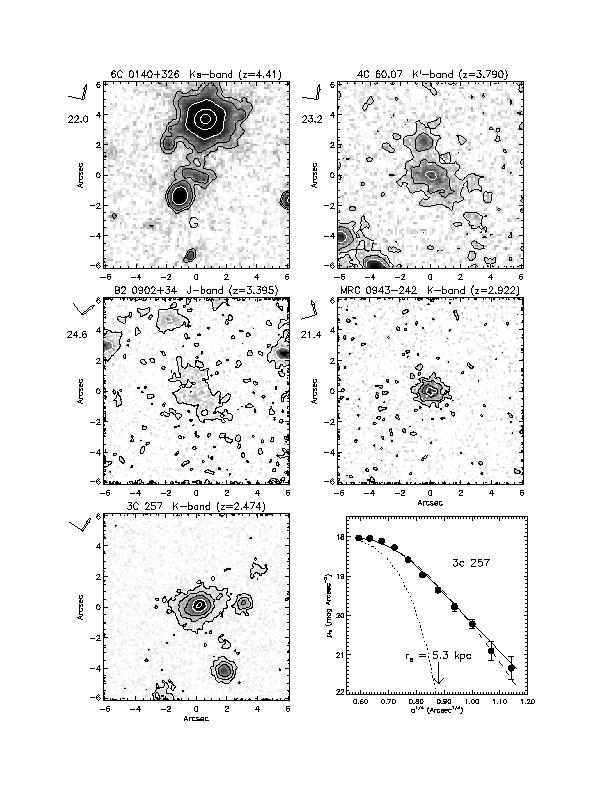}

\begin{figure}[h]
\caption{
Selected near--IR images of high redshift radio galaxies, presented in order of decreasing
redshift, and the surface brightness profile of 3C257.
\label{fig:nirc3c257}
}
\end{figure}

\vfill\eject


\hspace{2.0cm}

\noindent{This figure is available as 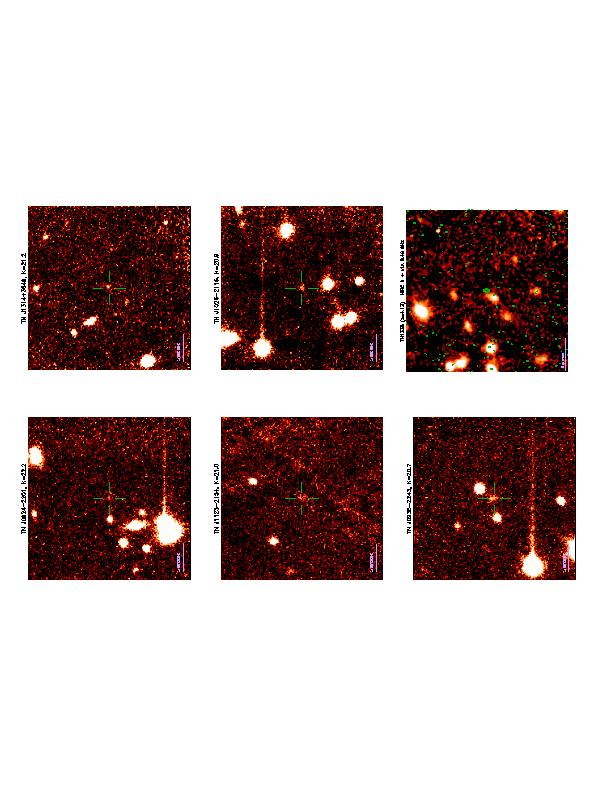}

\begin{figure}[h] 
\caption{ Some of our faintest Keck near-IR identifications of USS
sources.  Typical exposures with NIRC were $\sim 1$ hr. Spectroscopy
of one the faintest objects, \tnb, showed it to be at $z = 5.19$
(Section~\ref{sect:tn0924}).  
\label{fig:nircpanel} 
} 
\end{figure}

\vfill\eject


\hspace{2.0cm}

\noindent{This figure is available as 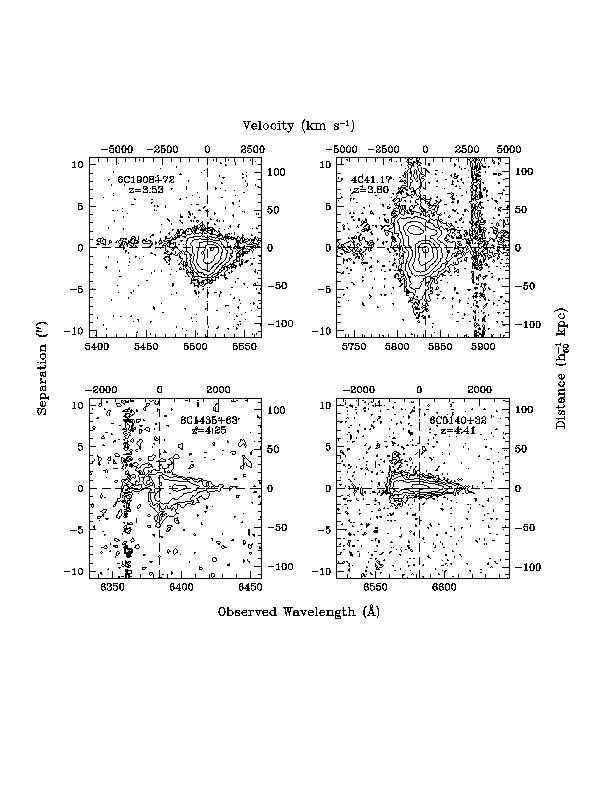}

\begin{figure}[h]
\caption{Ly-alpha emission profiles of four high redshift radio galaxies
(Figure and caption from Ref.~29). The triangular-shaped profiles result
from spatially extended absorption which is blue-shifted relative to the
systemic velocity. The zero velocity in all cases is determined either
from the HeII$\lambda$1640 line or the CIII{]}$\lambda$1909 line. Note
that the \lya\ emission from 4C41.17 extends over $>200$~kpc.
\label{fig:lya-abs}
}
\end{figure}

\vfill\eject


\vfill\eject


\vfill\eject


\noindent{This figure is available as 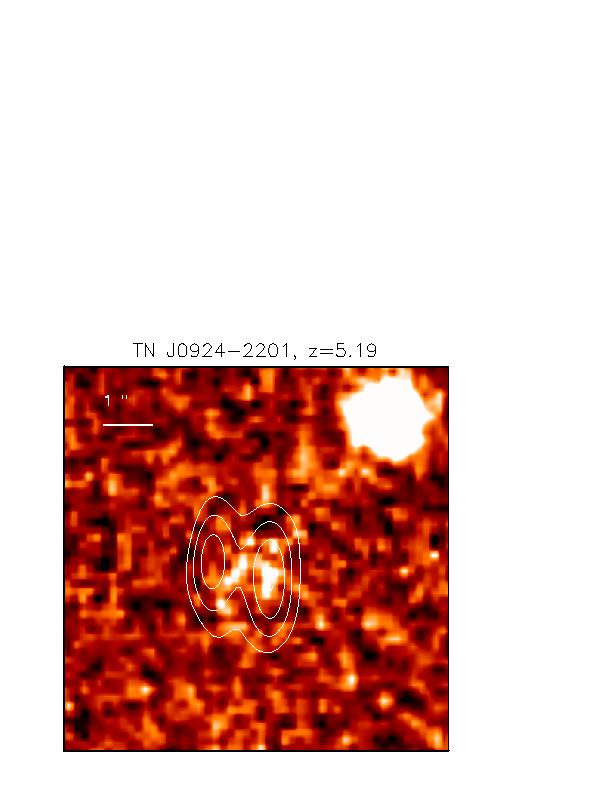}

\end{document}